\documentclass{appolb}
\usepackage{graphicx}
\usepackage{amsmath}
\usepackage{titlesec}
\allowdisplaybreaks

\begin{document}
\title{The real corrections to the Higgs impact factor at next-to-leading order with finite top mass
\thanks{Presented at ``Diffraction and Low-$x$ 2024'', Trabia (Palermo, Italy), September 8-14, 2024.}}

\author{Francesco Giovanni Celiberto
\address{Universidad de Alcalá (UAH), Departamento de Física y Matemáticas, Campus Universitario, Alcalá de Henares, E-28805, Madrid, Spain}
\\[3mm]
{Luigi Delle Rose, Gabriele Gatto\thanks{Speaker}, Alessandro Papa
\address{Dipartimento di Fisica, Università della Calabria, Arcavacata di Rende, I-87036, Cosenza, Italy}
\address{INFN, Gruppo Collegato di Cosenza, Arcavacata di Rende, I-87036, Cosenza, Italy}
}
\\[3mm]
Michael Fucilla
\address{Université Paris-Saclay, CNRS/IN2P3, IJCLab, F-91405, Orsay, France}
}

\maketitle
\begin{abstract}
This work presents the computation of real corrections to the impact factor for forward Higgs boson production, preserving the full dependence on the top-quark mass. The results are shown to align with the BFKL factorization framework, particularly in reproducing the expected rapidity divergence. Additionally, the subtraction of this divergence has been demonstrated using the appropriate counterterm within the BFKL scheme. In the infinite-top-mass limit, our findings reproduce the previously established result.
\end{abstract}

\section{Introduction}
The discovery of the Higgs boson at the Large Hadron Collider (LHC) marked the beginning of a new era for the precise examination of the Standard Model and the pursuit of phenomena beyond it. High-order calculations play a pivotal role in refining predictions for Higgs production under the standard collinear factorization. This approach relies on the convolution of universal, non-perturbative parton distribution functions (PDFs) with process-specific coefficient functions, which are computed perturbatively. Achieving high precision in these functions largely depends on incorporating QCD radiative corrections that extend beyond the leading order (LO).
This work delves into the \emph{semi-hard} regime, characterized by the scale hierarchy $\Lambda_{\rm QCD} \ll {Q_i} \ll \sqrt{s}$, where ${Q_i}$ represents a collection of process-dependent, hard scales, and $\sqrt{s}$ is the center-of-mass energy. In this regime, large energy logarithms emerge, and their resummation can be systematically achieved through the Balitsky--Fadin--Kuraev--Lipatov (BFKL) framework. This formalism provides a robust methodology for resumming such logarithms at both leading~\cite{Fadin:1975cb,Kuraev:1976ge,Kuraev:1977fs,Balitsky:1978ic} and next-to-leading~\cite{Fadin:1998py,Ciafaloni:1998gs} logarithmic levels of precision. For recent applications at the LHC, see~\cite{Celiberto:2022dyf,Celiberto:2022rfj,Celiberto:2024mab}. The BFKL cross sections can be expressed as convolutions of process-dependent impact factors and the Green's function, which has a universal character. An impact factor describes the transition from one of the initial-state particles to a specific object identified in the final state. This object is produced within the fragmentation region of the particle corresponding to the initial state.
Our work \cite{Celiberto:2024bbv} focuses on the calculation of real corrections to the NLO Higgs impact factor, arising from the emission of an additional parton in the fragmentation region where the Higgs is produced.
\begin{figure}[t]
\begin{picture}(430,70) % Mantieni l'altezza del box appropriata
\put(80,25){ \scalebox{5.0}{( }} % Sposta leggermente verso l'alto
\put(110,7){\includegraphics[scale=0.27]{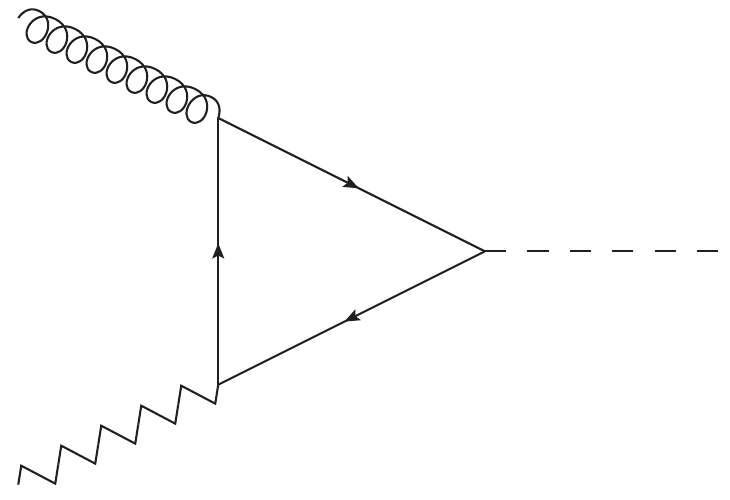}} % Aumenta la y
\put(128,70){\scalebox{0.8}{$k_1$}} % Spostato verso l'alto
\put(128,7){\scalebox{0.8}{$q$}} % Spostato leggermente verso l'alto
\put(182,47){\scalebox{0.8}{$p_H$}} % Sposta verso l'alto
\put(224,37){ $\times \hspace{0.15 cm} 2$} % Sposta verso l'alto
\put(230,25){ \scalebox{5.0}{ ) }} % Sposta leggermente verso l'alto
\end{picture} 
\caption{The two triangular-like diagrams that contribute to the Higgs impact factor at leading order (LO) are shown. The factor \( \times 2 \) accounts for the diagram where the direction of the fermion lines is reversed.}
\label{Fig:2DiagramsTriLO}
\end{figure}
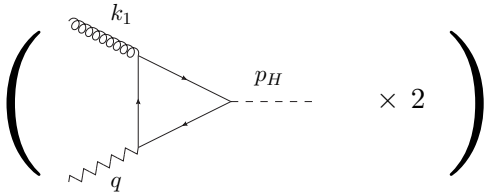
\section{LO computation}
At the level of hard scattering, the subprocess is initiated by a collinear gluon interacting with a $t$-channel Reggeon to produce the Higgs boson. To transition from the partonic subprocess to the hadronic, proton-initiated one, we employ the collinear factorization formula.
The gluon-initiated impact factor, differential with respect to the Higgs kinematic variables, is expressed as
\begin{align}
  \frac{ d \Phi_{PP}^{ \{ H \} (0) }(\vec q \; ) }{d x_H d^{2} \vec{p}_H } \!&=\! \int_{x_H}^1 \hspace{-0.2 cm} \frac{d z_H}{z_H} f_g \hspace{-0.1 cm}  \left( \hspace{-0.05 cm}  \frac{x_H}{z_H} \hspace{-0.05 cm}  \right)  \frac{ d \Phi_{gg}^{ \{ H \} (0) }(\vec q \; ) }{d z_H d^{2} \vec{p}_H }\nonumber\\ &= \frac{ |F_T \left( 0, -\vec{q}^{\; 2}, m_{H}^2  \right)|^2 \vec{q}^{\; 2} f_g (x_H) }{8 (1-\epsilon) \sqrt{N^2-1}} \delta^{(2)} \left( \vec{q} -\vec{p}_H \right),
  \label{Eq:LoImpactD4-2EHadro}
\end{align}
where \( f_g \) is the gluon distribution, \( d\Phi_{gg} \) is the differential impact factor for the production of a Higgs boson initiated by a gluon, $F_T$ is the form factor, \( \vec q \) represents the transverse momentum of the Reggeon, \( p_H \) is the transverse momentum of the Higgs, \( z_H \) denotes the longitudinal fraction of the Higgs with respect to the gluon, and \( x_H \) is the longitudinal fraction of the Higgs with respect to the proton. At leading order, since the initial gluon is collinear, the Reggeon momentum matches that of the Higgs, as evident from the delta function at the end of Eq.~(\ref{Eq:LoImpactD4-2EHadro}). Instead, the denominator \((1 - \epsilon)\) arises from averaging over the gluon polarizations in dimensional regularization.

\section{NLO computation}
\subsection{Impact factor for quark-initiated processes}
We begin with a discussion of the process in which an initial quark interacts with a Reggeized gluon to produce a Higgs boson and a final-state quark. The vertex is derived from two contributing diagrams, where a quark emits a gluon that subsequently interacts exactly as at leading order. These diagrams incorporate transverse ($F_T$) and longitudinal ($F_L$) form factors, with the latter appearing because both gluons producing the Higgs via the top-quark triangle are off-shell (see Ref.~\cite{Celiberto:2024bbv} for further details). Furthermore, we also note that in this case, due to the presence of an additional particle, the transverse momentum of the Higgs and that of the Reggeon are not necessarily equal, unlike at leading order.

The impact factor exhibits neither soft divergences nor rapidity divergences, ensuring a well-defined behavior in these regions. The only divergences present are collinear, which occur when $\vec{p}_q = \vec{q} - \vec{p}_H \to \vec{0}$. In the analysis presented in Ref.~\cite{Celiberto:2024bbv}, these collinear divergences have been shown to be consistent with the initial-state collinear divergences typically associated with the parton distribution functions. As such, these divergences are correctly accounted for and will cancel as expected in the context of physical cross-section calculations.
%_______________________________________________
\subsection{Impact factor for gluon-initiated processes}
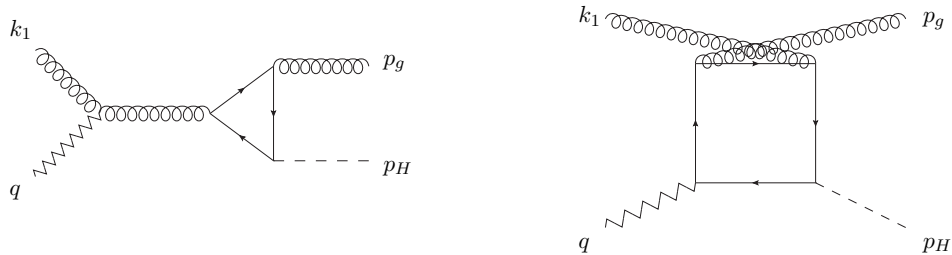
\begin{figure}
\begin{picture}(430,160)
\put(8,100){\includegraphics[scale=0.37]{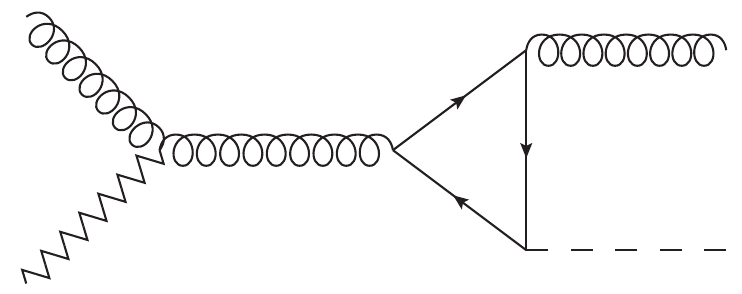}}
\put(3,155){\scalebox{0.8}{$k_1$}}
\put(3,95){\scalebox{0.8}{$q$}}
\put(143,144){\scalebox{0.8}{$p_g$}}
\put(143,104){\scalebox{0.8}{$p_H$}}
\put(223,80){\includegraphics[scale=0.35]{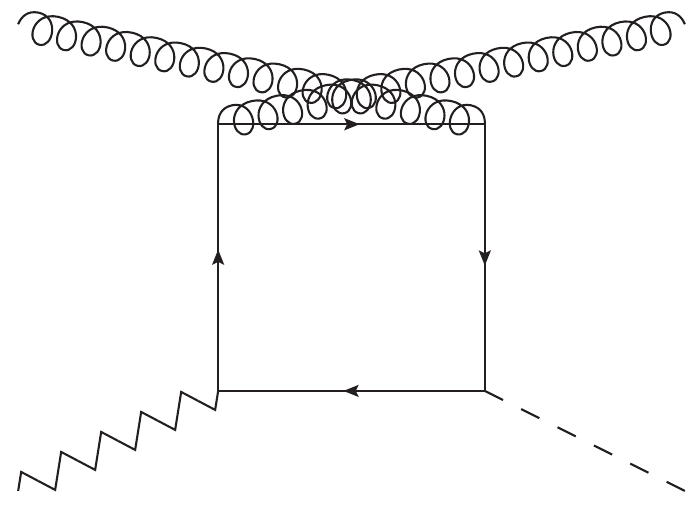}}
\put(216,160){\scalebox{0.8}{$k_1$}}
\put(216,75){\scalebox{0.8}{$q$}}
\put(345,160){\scalebox{0.8}{$p_g$}}
\put(345,75){\scalebox{0.8}{$p_H$}}
\end{picture} 
\vspace{-3cm} % Adatta il valore per ridurre lo spazio
\caption{An example of one of the six triangular-like diagrams and one of the six box-like diagrams contributing to the gluon-initiated contribution to the Higgs impact factor at NLO.}
\label{Fig:6Diagrams}
\end{figure}
This process involves the emission of a gluon, which can originate either from a gluon line or a quark line, followed by the production of a Higgs boson. The contributions to this process can be classified into two distinct categories: those arising from triangular-like diagrams and those from box-like diagrams (see Fig.~\ref{Fig:6Diagrams}). The boxes have also been expressed in Ref.~\cite{Celiberto:2024bbv} in terms of certain form factors. The singularities can be classified as:
\begin{itemize}
\item[\textbullet]\textbf{Collinear singularities}: These arise when $\vec{p}_g = \vec{q} - \vec{p}_H \to \vec{0}$, with the longitudinal momentum fraction $z_g = 1 - z_H$ fixed. Such divergences are shown to be consistent with those expected from the initial-state gluon PDF. As such, they are absorbed into the renormalization of the gluon PDF, ensuring the proper cancellation.
    \item[\textbullet]\textbf{Soft singularities}: These occur when $\vec{p}_g = \vec{q} - \vec{p}_H = (1-z_H) \vec{u}$, with $z_g \to 0$. The gluon is emitted with negligible energy, leading to a divergence in the real-emission phase space. A direct cancellation occurs between the real and virtual contributions within the same phase space region, ensuring the infrared finiteness of the total impact factor.
\item[\textbullet]\textbf{Rapidity singularities}: emerge in the limit $z_H \to 1$. In the high-rapidity limit, the impact factor is expressed as:
\begin{equation*}
   \frac{d \Phi_{gg}^{ \{H g \}} (z_H, \vec{p}_H, \vec{q} ; s_0)}{d z_H d^2 \vec{p}_H} \bigg|_{z_H \rightarrow 1} 
\end{equation*}
\begin{equation}
    = \frac{g^2 |F_T ( 0, -\vec{p}_H^{\; 2}, m_{H}^2  )|^2 N}{4 (1-\epsilon) \sqrt{N^2-1} (2 \pi)^{D-1} } \frac{\vec{q}^{\; 2}}{(\vec{q}- \vec{p}_H)^2} \frac{1}{(1-z_H)} \theta \left( s_{\Lambda} - \frac{(\vec{q}-\vec{p}_H)^2}{(1-z_H)} \right) \; .
\label{Eq:HighRapidityLimitIF}
\end{equation}
In this expression, the parameter $s_\Lambda$ acts as a regulator to manage the divergent behavior. This divergence is removed through the introduction of a BFKL counter-term, which depends on $s_\Lambda$ and cancels the rapidity-dependent contributions.
\end{itemize}
At the end, it is demonstrated that the impact factor remains consistent with its gauge-invariant definition, utilizing the \( m_t \to \infty \) expansion up to next-to-next-to-leading order (NNLO).

\section{Summary and conclusions}
We calculated the real corrections to the next-to-leading order Higgs impact factor, arising from the emission of an additional parton in the Higgs production fragmentation region. Our work incorporates a finite top-quark mass in the Higgs impact factor calculation, advancing beyond the infinite-mass approximation~\cite{Nefedov:2019mrg, Hentschinski:2020tbi, Celiberto:2022fgx}. We confirmed gauge invariance and absence of rapidity divergences, with consistent indications of proper infrared behavior.
The next and conclusive step in this line of research will involve the calculation of virtual corrections, an aspect that will be addressed in a forthcoming publication. The interest in virtual corrections is twofold. From one side, once completed, virtual corrections will enable more precise predictions for forward Higgs production processes at the LHC and future colliders~\cite{Bonvini:2018ixe,Celiberto:2020tmb,Andersen:2018kjg}, exploring new kinematic regions with next-to-leading logarithmic resummation. From a more formal perspective, the calculation of virtual corrections is important to confirm the consistency of the results obtained adopting the infinite-top-mass approximation with the gluon Reggeization. This latter has been shown to hold at one-loop level in a completely non-trivial way, due to the presence of the effective non-renormalizable Higgs-gluon coupling~\cite{Fucilla:2024cpf}. Calculating the virtual corrections will provide an all-order proof. 
Virtual corrections are essential for completing the NLO analysis and achieving a fully consistent description of the Higgs impact factor in this framework.

\section{Acknowledgments}
We thank Maxim A. Nefedov, Samuel Wallon, Renaud Boussarie, Lech Szymanowski, Fulvio Piccinini, Vittorio Del Duca, Carlo Oleari, Carl R. Schmidt and Dieter Zeppenfeld. The work of F.\allowbreak G.\allowbreak C.\ is supported by the Atracción de Talento Grant No. 2022-T1/TIC-24176 (Madrid, Spain); L.\allowbreak D.\allowbreak R., G.\allowbreak G., and A.\allowbreak P.\ by the INFN/QFT@COLLIDERS Project (Italy); and M.\allowbreak F.\ by ANR under contract ANR-17-CE31-0019. M.\allowbreak F.\ acknowledges support from the Italian Foundation “Angelo Della Riccia”. All pictures in this work have been drawn using JaxoDraw.

\end{document}